\DeclareUnicodeCharacter{FF0C}{,} 
\documentclass[sigconf]{acmart}

\usepackage{pifont}
\usepackage{multirow}
\usepackage{lipsum}

\usepackage{amssymb}
\usepackage{enumitem}
\usepackage{threeparttable}
\usepackage{xcolor}
\usepackage{array}
\usepackage{makecell}    
\usepackage{colortbl}    
\usepackage{amsmath}   
\usepackage{blindtext}

\newcommand{\header}[1]{\noindent \textbf{#1.}}

\newcommand{\dataset}{\textsc{FinTMMBench}}
\newcommand{\method}{\textsc{TMMHybridRAG}}

\AtBeginDocument{%
  }

\author{Fengbin Zhu}
\authornote{Equal Contribution}
\affiliation{
  \institution{National University of Singapore}
    \country{Singapore}
}
\email{zhfengbin@gmail.com}

\author{Junfeng Li}
\authornotemark[1]
\affiliation{
  \institution{National University of Singapore}
  \country{Singapore}
}
\email{lijunfeng@u.nus.edu}

\author{Liangming Pan}
\authornote{Corresponding Author}
\affiliation{
  \institution{Peking University}
  \country{China}
}
\email{peterpan10211020@gmail.com}

\author{Wenjie Wang}
\affiliation{
  \institution{University of Science and Technology of China}
  \country{China}
}
\email{wenjiewang96@gmail.com}

\author{Fuli Feng}
\affiliation{
  \institution{University of Science and Technology of China}
\country{China}
}
\email{fulifeng93@gmail.com}

\author{Chao Wang}
\affiliation{
  \institution{6Estates Pte Ltd}
  \country{Singapore}
}
\email{wangchao@6estates.com}

\author{Huanbo Luan}
\affiliation{
  \institution{6Estates Pte Ltd}
  \country{Singapore}
}
\email{luanhuanbo@6estates.com}

\author{Tat-Seng Chua}
\affiliation{
  \institution{National University of Singapore}
  \country{Singapore}
}
\email{chuats@comp.nus.edu.sg}

\setcopyright{acmlicensed}
\copyrightyear{2018}
\acmYear{2018}
\acmDOI{XXXXXXX.XXXXXXX}
\acmConference[Conference acronym 'XX]{Make sure to enter the correct
  conference title from your rights confirmation emai}{June 03--05,
  2018}{Woodstock, NY}
\acmISBN{978-1-4503-XXXX-X/18/06}

\renewcommand{\shortauthors}{Tor et al.}

\begin{document}

\title{Towards Temporal-Aware Multi-Modal Retrieval Augmented Generation in Finance}





\begin{abstract}
Finance decision-making often relies on in-depth data analysis across various data sources, including financial tables, news articles, stock prices, etc. 
In this work, we introduce \dataset, the first comprehensive benchmark for evaluating temporal-aware multi-modal Retrieval-Augmented Generation (RAG) systems in finance. 
Built from heterologous data of NASDAQ 100 companies, 
\dataset~offers three significant advantages.  
1) \emph{Multi-modal Corpus}: It encompasses a hybrid of financial tables, news articles, daily stock prices, and visual technical charts as the corpus.
2) \emph{Temporal-aware Questions}: Each question requires the retrieval and interpretation of its relevant data over a specific time period, including daily, weekly, monthly, quarterly, and annual periods.  
3) \emph{Diverse Financial Analysis Tasks}: The questions involve 10 different financial analysis tasks designed by domain experts, including information extraction, trend analysis, sentiment analysis and event detection, etc.  
We further propose a novel \method~method, which first leverages a multi-modal LLM to convert data from other modalities (e.g., tabular, visual and time-series data) into textual format and then incorporates temporal information in each node when constructing graphs and dense indexes. 
Its effectiveness has been validated in extensive experiments, but notable gaps remain, highlighting the challenges presented by our \dataset. 
The benchmark and source code will be made publicly available\footnote{https://github.com/lijunfeng99/FinTMMBench}.
\end{abstract}



\received{20 February 2007}
\received[revised]{12 March 2009}
\received[accepted]{5 June 2009}

\maketitle


\section{Introduction}

\begin{figure}[t]
    \centering
    \setlength{\abovecaptionskip}{-0.3pt}
    \setlength{\belowcaptionskip}{-0.3pt}
    \includegraphics[width=\linewidth]{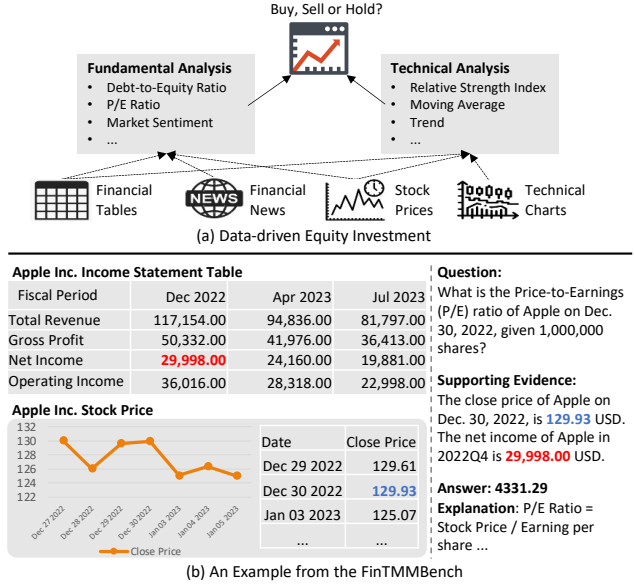}
    \caption{ (a) Illustration of financial analysis for decision-making. (b) An example from \dataset.}
    \label{fig:FinancialAnalysis}
    \vspace{-0.4cm}
\end{figure}

Financial analysis is fundamental to modern finance, supporting applications such as equity investment~\cite{fama1992cross}, portfolio optimization~\cite{markowitz1991foundations}, and risk management~\cite{power2004risk}. 
Effective decision-making in these areas requires synthesizing up-to-date information from diverse modalities, including structured tables, unstructured text, time-series data, and visual charts, as illustrated in \autoref{fig:FinancialAnalysis} (a).

\begin{figure*}[t]
    \centering
    \setlength{\abovecaptionskip}{-0.0pt}
    \setlength{\belowcaptionskip}{-0.0pt}
    \resizebox{0.95\linewidth}{!}{\includegraphics{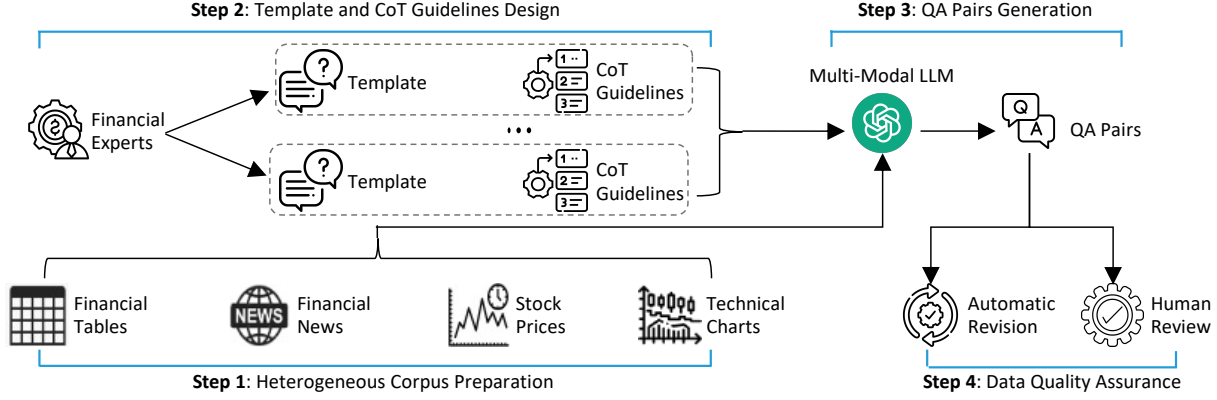}}
    \caption{An overall pipeline for constructing \dataset.}
    \label{fig:Dataset_process}
    \vspace{-0.4cm}
\end{figure*}



Recently, Retrieval-Augmented Generation (RAG) systems have been increasingly explored in financial analysis~\cite{wang2024omnieval,li2024alphafin}.
Current financial benchmarks for evaluating RAG systems include FinTextQA~\cite{chen2024fintextqa}, AlphaFin~\cite{li2024alphafin},  OmniEval~\cite{wang2024omnieval}, and FinanceBench \cite{islam2023financebench}. 
However, these datasets offer limited data modalities, potentially harming the validity of evaluation. 
Specifically, FinTextQA and OmniEval are restricted to textual data, whereas AlphaFin covers textual and time-series data, and FinanceBench combines textual and visual data.
In addition, they often fail to adequately incorporate temporal information in their task design, which is critical for assessing whether RAG systems can accurately retrieve and process financial data within specific time periods.
Although AlphaFin introduces some temporal questions, they are solely centered on time-series data.
Their narrow focus restricts their ability to comprehensively evaluate RAG systems in handling \textbf{temporal-aware queries} over heterogeneous data across \textbf{different modalities}.


To address these gaps,we introduce \textbf{\dataset}, a financial benchmark for RAG evaluation in equity investment, integrating diverse data types for comprehensive analysis. 
As shown in \autoref{fig:FinancialAnalysis} (a), financial tables and news articles are simultaneously used for calculating key financial ratios and assessing market sentiment in fundamental analysis, and stock prices and technical charts are both required for calculating moving averages and identifying trends in technical analysis.
Furthermore, equity analysis often involves temporal-aware queries, which require precise identification of time-specific information(e.g. year, month). 
For example, as shown in \autoref{fig:FinancialAnalysis} (b), answering ``\emph{What is the Price-to-Earnings (P/E) ratio of Apple on Dec 30, 2022, given 1,000,000 shares?}'' requires extracting data for ``Dec 30, 2022'' from tables and stock prices, highlighting the need for temporal awareness.


To construct \dataset, we collect 2022 financial data for all NASDAQ-100 companies across four modalities. Working with financial experts, we use a template-based approach to automatically generate QA pairs, reflecting real-world analysis needs. About 100 templates with Chain-of-Thought (CoT) guideline cover 10 financial tasks, such as information extraction, trend analysis and event detection. Automatic revision and human review further enhance data quality. In total, \dataset~contains $5,676$ high-quality QA pairs and $36,100$ raw data items.


Existing RAG methods, such as GraphRAG~\cite{edge2024local} and LightRAG~\cite{guo2024lightrag}, tend to struggle with answering the temporal-aware questions across multi-modal financial data in our \dataset, as shown in \autoref{tab:OverallPerformanceComparison}.  
To address the challenge in \dataset, we propose a novel \textbf{\method}~method by combining dense retrieval and graph retrieval techniques. 
First, \method~extracts entities and their relations from each financial news article and employs an LLM to generate descriptions for each entity and relation.
For non-textual data, \method~regards each table, daily stock price record, and chart as a distinct entity and utilizes an advanced multi-modal LLM to generate a textual summary for each, which serves as the entity's description.
Further, \method~integrates temporal information into every entity and relation as the properties to construct dense vectors and graphs.
During prediction, given a question, all retrieved entities and relations from both dense vectors and graphs, along with their raw data, are fed into a multi-modal LLM to infer the answer.
Extensive experiments show that our \method~ method significantly outperforms all compared methods across all evaluation metrics. 
However, its F1 score remains relatively low at 31.41, highlighting the substantial challenges presented in \dataset~and underscoring the need for more advanced RAG methods.

In summary, our major contributions are threefold:
    1) To the best of our knowledge, we are the first to investigate temporal-aware multi-modal RAG in the financial domain, addressing a critical real-world need in financial analysis. 
    2) We introduce a new benchmark, \dataset, specially designed to evaluate temporal-aware multi-modal RAG systems in finance. \dataset~comprises 5,676 temporal-aware questions that require information from four distinct modalities, i.e. financial tables, news articles, daily stock prices, and visual technical charts, to be answered.
    3) To tackle the challenges in \dataset, we propose \method, a novel temporal-aware multi-modal RAG method that integrates dense and graph retrieval techniques.
    Experiments demonstrate that our \method~beats all compared methods, serving as a strong baseline on \dataset.
    

\begin{figure}
    \centering
\setlength{\abovecaptionskip}{-0.4pt}
\setlength{\belowcaptionskip}{-6pt}
    \includegraphics[width=0.85\linewidth]{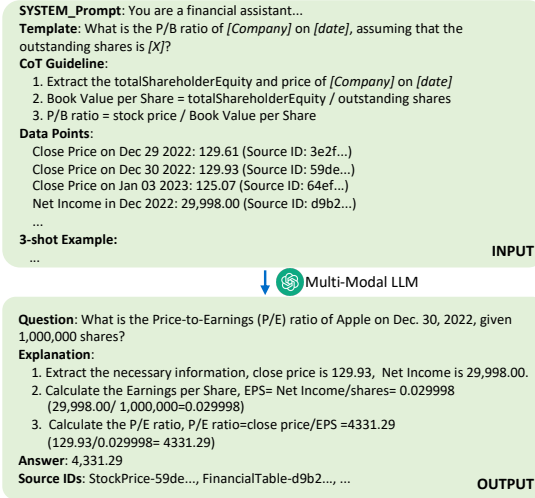}
    \caption{An example for QA pair generation.
    }
    \label{fig:QAGen}
    \vspace{-0.5cm}
\end{figure}

\section{Proposed \dataset}
Our \dataset~is constructed following a template-guided generation pipeline, as shown in~\autoref{fig:Dataset_process}.

\subsection{Heterogeneous Corpus Preparation}
To construct \dataset, we collect financial data of the \textit{NASDAQ-100} companies in 2022, which include four types as below. 

\begin{itemize}[leftmargin=*]
\item \textbf{Financial Tables}:.For each company, we collect 12 quarterly and 3 annual financial tables from 2022 via public APIs\footnote{https://www.alphavantage.co/}, totaling 1,500 financial tables.
\item \textbf{News Articles}: We gather over $70,000$ Reuters financial news articles from 2021--2022, then filter for strong relevance to \textit{NASDAQ-100} companies, resulting in about $3,100$ articles.
\item \textbf{Daily Stock Prices}: For each company, we collect 252 daily records (high, low, open, close, volume) for 2022, totaling $25,200$ records.
In total, we obtain $25,200$ records for all the companies.
\item \textbf{Visual Technical Charts}: Weekly and monthly candlestick charts are generated from the daily stock price data.
\end{itemize}


All data are standardized into JSON files, each storing a granular data point(e.g., a news article about a company), ensuring consistency and seamless integration across modalities.

\subsection{Template and CoT Guidelines Design}

Considering the high cost of human annotation, we design diverse question templates and corresponding CoT guidelines, to guide multi-modal LLMs to generate high-quality QA pairs automatically.
Specifically, we collaborate with financial experts to curate a set of approximately $100$ different question templates, which cover various financial tasks including:
\begin{itemize}[leftmargin=*] 
    \item \textbf{Information Extraction (IE)}: The question requires querying specific information (e.g.,  total revenue and net income) from the financial corpus.
    \item \textbf{Arithmetic Calculation (AC)}: The question requires deriving an indicator using a given formula based on  relevant information.
    \item \textbf{Trend Analysis (TA)}: The question requires analyzing the trend of an indicator over time.
    \item \textbf{Logical Reasoning (LR)}: The question requires logical reasoning to infer the answer.
    \item \textbf{Sentiment Classification (SC)}: The question requires analyzing the sentiment polarity of a news article relevant to a specific company's aspect (e.g., product and service). 
    \item \textbf{Event Detection (ED)}: The question requires identifying the events mentioned in a news article.
    \item \textbf{Counterfactual Reasoning (CR)}: The question requires counterfactual reasoning to answer.
    \item \textbf{Comparison (CP)}: The question requires comparing indicators across different companies to obtain the answer. 
    \item \textbf{Sorting (ST)}: The question requires sorting indicators to infer the answer. 
    \item \textbf{Counting (CT)}: The question requires counting the number of data points to infer the answer.
\end{itemize}

All questions are temporal-aware, requiring information from specific periods (e.g., day, month), and may encompass multiple financial tasks.
Detailed CoT guidelines for each template encourage step-by-step reasoning, reducing inconsistencies in generated QA pairs and improving the quality of the dataset.

\subsection{QA Pair Generation}

\begin{table}[t]
\centering
\caption{Statistics of \dataset.}
\setlength{\abovecaptionskip}{-0.6pt}
\setlength{\belowcaptionskip}{-0.8cm}

\begin{tabular}{lr}
\hline
\textbf{Statistic} & \textbf{Number} \\
\hline
    Total Number of Companies & 100 \\
\hline
    Total Number of Raw Data & 36,100 \\
    \:  \# Financial Tables & 1,500 \\
    \:  \# News Articles & 3,133 \\
    \:  \# Daily Stock Price & 25,200 \\
    \:  \# Visual Technical Charts & 6,267 \\
\hline
    Total Number of Questions & 5,676 \\
    Avg. Number of Question per Company & 56.76 \\
    Avg. Number of Words per Question & 18.88 \\
    Avg. Number of Words per Answer & 6.87 \\
\hline
\end{tabular}%

\label{tab:dataset_summary}
\vspace{-0.55cm}
\end{table}

We employ GPT-4o-mini as the multi-modal LLM for QA pair generation.  
As shown in \autoref{fig:QAGen}, the multi-modal LLM receives three key inputs: 1) a question template, 2) a CoT guideline, and  3) some data points of daily stock prices.
Only data points relevant to each question template (e.g., news articles for event detection) are provided as input to the multi-modal LLM.
This allows the multi-modal LLM to focus on essential information for QA generation. We prompt it to generate a question, step-by-step reasoning, the final answer, and the IDs of referenced data points. Few-shot prompting is used to further improve QA quality.

\subsection{Data Quality Assurance}
\noindent \textbf{Automatic Revision}. We preform automatic revision and human review to ensure the data quality of \dataset.
Specifically, we develop a script to automatically check and revise the generated QA pairs based on predefined rules.
To name a few, the IDs of the referred data points must be correct; the equations in each reasoning step must maintain equality between the left and right sides; the answer inferred based on all reasoning steps must be consistent with the final answer.

\noindent \textbf{Human Review}. After each round of automatic revision, we randomly select a set of samples based on the distribution of financial tasks and have two domain experts evaluate their accuracy, documenting any issues, which then inform the next round of automatic revision. 
The verification results are then reviewed by a third expert for additional validation. 
We repeat this iterative revision-review process until the verification accuracy  exceeds $85\%$ and the inter-annotator agreement between the two experts reaches $85\%$.

\subsection{Dataset Analysis}

As shown in \autoref{tab:dataset_summary}, \dataset~consists of $34,815$ raw data entries from \textit{NASDAQ-100} companies across four modalities, including $1,500$ financial tables, $3,133$ news articles, $25,200$ daily stock price records, and $6,267$ visual technical charts.
A total of $5,676$ QA pairs are generated based on these raw data and CoT templates, with average length of the questions is $18.88$ words, and the average length of the answers is $6.87$ words.

\autoref{tab:task_distribution} shows the distribution of financial tasks across data modalities. The questions in \dataset~span a wide range of financial tasks, enabling comprehensive evaluation of RAG systems on heterogeneous financial data.

\begin{table}[t]
\centering
\footnotesize
\caption{Financial task distribution across different modalities in \dataset.}
\setlength{\abovecaptionskip}{-0.3pt}
\setlength{\belowcaptionskip}{-20pt}
\setlength{\tabcolsep}{2.2mm}
\begin{tabular}{lrrrrr}
\hline
\bf FA Task & \bf Table & \bf News & \bf Price & \bf Chart & \bf Hybrid \\
\hline
    Information Extraction & 1,950 & 0     & 1,315 & 0     & 416 \\
    Arithmetic Calculation & 1,494 & 0     & 1,112 & 0     & 416 \\
    Trend Analysis & 575   & 0     & 489   & 421   & 0 \\
    Logical Reasoning & 661   & 0     & 121   & 0     & 164 \\
    Sentiment Classification & 0     & 977   & 0     & 0     & 0 \\
    Event Detection & 0     & 597   & 0     & 0     & 0 \\
    Counterfactual Reasoning & 778   & 0     & 539   & 0     & 416 \\
    Comparison & 474   & 0     & 673   & 0     & 302 \\
    Sorting & 560   & 0     & 166   & 0     & 95 \\
    Counting & 123   & 0     & 0     & 0     & 0 \\
\hline
\end{tabular}
\label{tab:task_distribution}

\vspace{-0.5cm}
\end{table}
\subsection{Comparison with Other Benchmarks}
We further provide a comparison of our \dataset~with existing financial QA datasets to stress its merits, as shown in  \autoref{tab:financial_qa_datasets}.
It can be seen that most existing financial QA datasets are not open-domain, except for FinTextQA \cite{chen2024fintextqa}, FinanceBench \cite{islam2023financebench}, AlphaFin~\cite{li2024alphafin} and OmniEval~\cite{wang2024omnieval}. 
With the exceptions of TempQuestions \cite{jia2018tempquestions} and AlphaFin \cite{li2024alphafin}, few of them are designed to address temporal-aware questions.
In addition, existing datasets are mostly restricted to specific modalities, such as textual data only (e.g., FinTextQA~\cite{chen2024fintextqa}), time-series data only (e.g., TempQuestions~\cite{jia2018tempquestions}),  both tabular and textual data (e.g., TAT-QA~\cite{zhu2021tat}),  or tabular, textual, and visual data (e.g., MultiModalQA~\cite{talmormultimodalqa}).
Compared with them, our \dataset~is designed to evaluate RAG systems in answering temporal-aware questions across a multi-modal corpus, encompassing tabular, textual, time-series, and visual data.

\begin{table}[t]
\centering
\caption{Comparison between our \dataset~with other Financial QA Datasets.}
\setlength{\abovecaptionskip}{-20pt}
\setlength{\belowcaptionskip}{-50pt}
\setlength{\tabcolsep}{0.6mm}
\footnotesize
\begin{tabular}{l |c | c| c c c c} 
\hline
\multirow{2}{*}{\textbf{Dataset}} &\multirow{2}{*}{\textbf{RAG}} & \multirow{2}{*}{\textbf{\makecell[l]{Temporal \\ Question}}} & \multicolumn{4}{c}{\textbf{Corpus Modality}}\\ 
& & & \bf Tabular & \bf Textual & \bf Time-Series & \bf Visual \\ \hline
FiQA-SA \cite{maia201818}       & \ding{55} & \ding{55} & \ding{55} & \checkmark & \ding{55} & \ding{55}\\ 
FPB \cite{malo2014good}         & \ding{55} & \ding{55} & \ding{55} & \checkmark & \ding{55} & \ding{55}\\ 
TAT-QA \cite{zhu2021tat}    & \ding{55} & \ding{55} & \checkmark & \checkmark & \ding{55} & \ding{55}\\ 
TAT-HQA \cite{li2022learning}    & \ding{55} & \ding{55} & \checkmark & \checkmark & \ding{55} & \ding{55}\\ 
FinQA \cite{chen2021finqa}   & \ding{55} & \ding{55} & \checkmark & \checkmark & \ding{55} & \ding{55}\\ 
MultiHiertt \cite{zhao2022multihiertt}   & \ding{55} & \ding{55} & \checkmark & \checkmark & \ding{55} & \ding{55}\\ 
FinBen~\cite{xie2024finben}  & \ding{55} & \ding{55} & \ding{55} & \checkmark & \ding{55} & \ding{55}\\ 
TAT-DQA \cite{zhu2022towards}   & \ding{55} & \ding{55} & \checkmark & \checkmark & \ding{55} & \checkmark\\ 
MultiModalQA \cite{talmormultimodalqa} & \ding{55} & \ding{55} & \checkmark & \checkmark & \ding{55} & \checkmark\\ 
TempQuestions \cite{jia2018tempquestions} & \ding{55} & \checkmark & \ding{55} & \checkmark & \checkmark & \ding{55}\\
AlphaFin \cite{li2024alphafin}  & \checkmark & \checkmark & \ding{55} & \checkmark & \checkmark & \ding{55}\\ 
FinTextQA \cite{chen2024fintextqa} & \checkmark & \ding{55} & \ding{55} & \checkmark & \ding{55} & \ding{55}\\ 

OmniEval \cite{wang2024omnieval}   & \checkmark & \ding{55} & \ding{55} & \checkmark & \ding{55} & \ding{55}\\ 

FinanceBench \cite{islam2023financebench}  & \checkmark & \ding{55} & \ding{55} & \checkmark & \ding{55} & \checkmark \\ 
\hline
\dataset & \checkmark & \checkmark & \checkmark & \checkmark & \checkmark & \checkmark\\

\hline
\end{tabular}
\label{tab:financial_qa_datasets}
\vspace{-0.5cm}
\end{table}

\begin{figure*}
    \centering
    
\setlength{\abovecaptionskip}{0.3pt}
\setlength{\belowcaptionskip}{0.3pt}     
\includegraphics[width=0.96\linewidth]{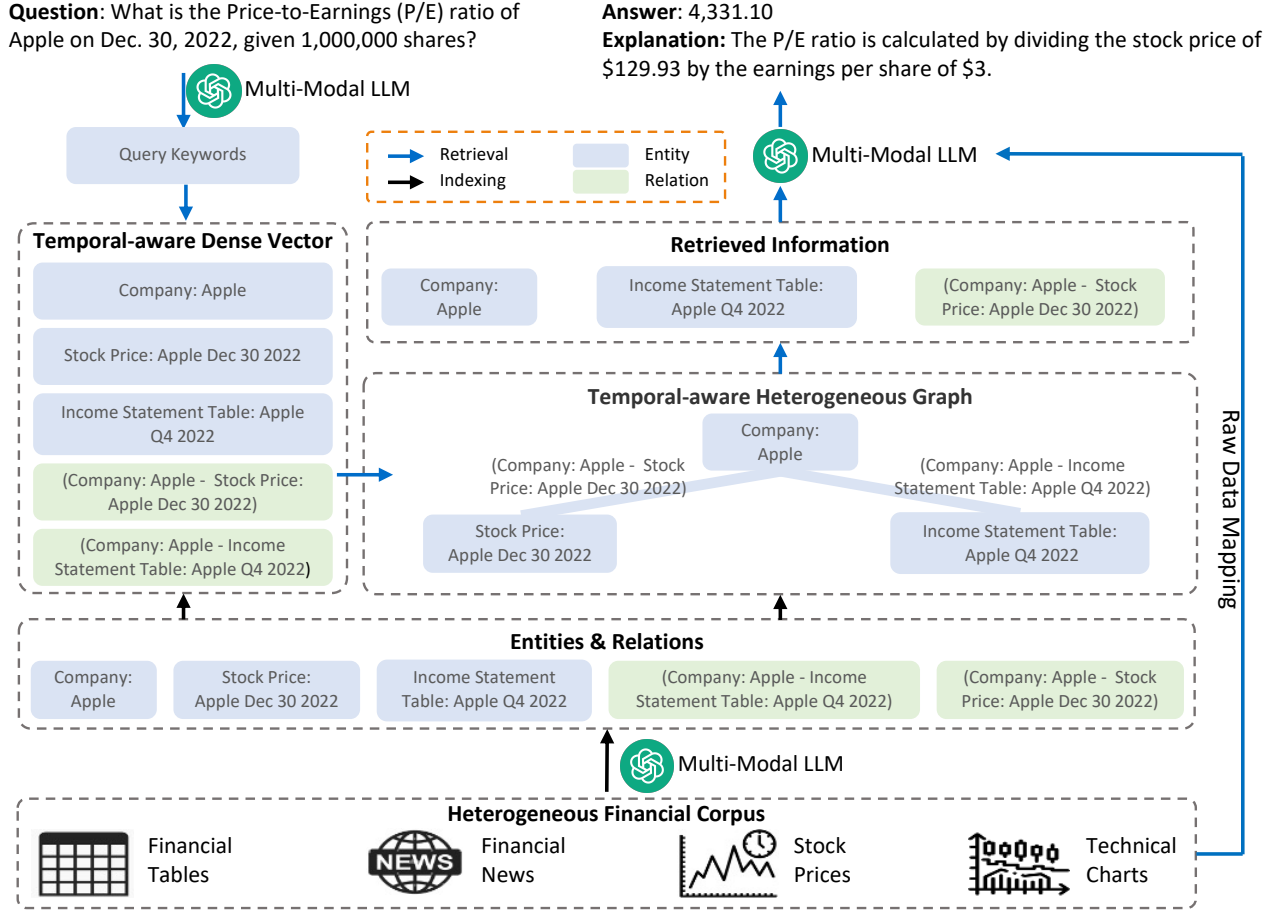}
    \caption{Illustration of proposed \method, a novel Temporal-Aware Multi-Modal RAG method.}
    \label{fig:method}
    \vspace{-0.5cm}
\end{figure*}

\section{Proposed~\method~Method}
To address the temporal-aware questions over heterogeneous financial data in \dataset,  we propose a novel RAG method \method, which combines the dense and graph retrieval techniques, as shown in \autoref{fig:method}.

\subsection{Preprocessing}
We generate textual descriptions for all non-textual data and then identify entities and their relationships across different modalities as preprocessing. In particular,

\begin{itemize}[leftmargin=*]
\item \textbf{Financial Tables:} Each financial table is treated as an entity with the temporal information determined by the period involved in the table, and its name involves the company name, table name, and the period described in this table. 
A summary of table is generated by an LLM, serving as the entity's description.



\item \textbf{News Articles}: 
The enenties and relationships with their descriptions are directly extracted from each news article using an LLM.
The temporal information of an entity or relationship is the publication date of the news article.


\item \textbf{Daily Stock Prices:} Each daily stock price record is treated as a unique entity, named with the stock symbol and date. An LLM generates a description for each record, and the date serves as its temporal information. We link records from consecutive business days for the same company to capture the temporal information.

\item \textbf{Visual Technical Charts:} Each chart is regarded as an entity, with its name incorporating metadata like the company name, and the time period represented in the chart.
Then, we utilize a multi-modal LLM to generate a concise summary for each chart, which serves as the entity's description. 
The period depicted in the chart serves as the temporal information for the entity.

\item \textbf{Cross-Modality Relationships:}
Cross-modality relationships play a critical role in unifying the diverse data sources within the temporal knowledge graph. 
Specifically, we employ a multi-modal LLM to automatically establish relationships across different modalities by providing it the contextual information about the entities, including their names, associated metadata, and textual descriptions. With this information, the MLLM infers and generates cross-modality relationships by identifying logical connections between the entities.
\end{itemize}

\subsection{Indexing}

\header{Temporal-aware Dense Vectors} First, \method~encodes each entity and relationship with its temporal information to generate a dense vector using OpenAI embedding models (i.e., OpenAI text-embedding-3-small).
Then, we store all obtained dense vectors in a vector database for further usage in the retrieval phase. 
Embedding temporal information directly into the hidden representations allows for the retrieval of relevant entities and relationships based on their associated date period.

\header{Temporal-aware Heterogeneous Graph}
Knowledge graphs~\cite{google2012knowledgegraph} are powerful tools for representing relationships between diverse entities. 
\method~builds a knowledge graph with extracted entities, e.g. company, person and location, and their relations with an online LLM (i.e., GPT-4o-mini).
Given the importance of temporal information in the finance domain, each entity and relationship is designed to store its corresponding temporal information as one of its properties.
Additionally, each entity and relationship includes a \textit{textual description property} and a \textit{Source ID} attribute that facilitates raw data mapping during the generation phase.

\subsection{Retrieval}
We integrate dense and graph retrieval for enhanced effectiveness.

\header{Keywords Identification and Expansion} Given a question, we first use an LLM to extract and expand relevant keywords, following an approach similar to LightRAG~\cite{guo2024lightrag}. These keywords, including both entity and relationship names, are utilized to retrieve relevant entities and relationships from the dense vectors and graph.

\header{Dense Retrieval} 
We encode each query keyword into a dense vector and retrieve the top $K$ vectors from the vector database.
Each dense vector represents an entity or a relation.

\header{Graph Retrieval}  
First, we aggregate all query keywords along with entity and relationship names obtained from dense retrieval.
We then use these combined keywords to apply graph retrieval, searching for associated entities and relationships within the graph.
Finally, all retrieved entities and relationships from both the vector and the graph are utilized to generate the answer in the next step.

\subsection{Generation}
With the retrieved entities and relations, we leverage a multi-modal LLM to generate the final answer. 


\header{Raw Data Mapping}
First, we gather the raw data from different modalities linked to the retrieved entities and relationships based on the source IDs.
Although we generate a textual description for each entity and relation, some crucial information or metrics may be inadvertently lost without the raw data.
By providing original data sources, we ensure that any analysis conducted is based on the correct and complete information.





\header{Answer Generation} A multi-modal LLM is utilized to generate the final answer, taking as input the question, the retrieved entities and relationships along with their temporal properties and textual descriptions, and the corresponding raw data.  
The multi-modal LLM is instructed to output the intermediate reasoning steps and the final answer based on the multi-modal inputs.

\section{Experiments}


\subsection{Experimental Settings}

\header{Compared Methods}
We employ three experimental settings.
1) \emph{No Retrieval}: No data retrieval is applied, and only the question itself is fed into a multi-modal LLM to infer the answer;
GPT-4o-mini is adopted in this setting.
2) \emph{No Visual}: The retrieved tables, news, stock prices and textual description of charts are fed into a multi-modal LLM. BM25~\cite{robertson2009probabilistic}, Naive RAG~\cite{gao2023retrieval}, GraphRAG~\cite{edge2024local}, LightRAG~\cite{guo2024lightrag}, and BGE-Text~\cite{bge_embedding} are applied in this setting. 
3) \emph{All}: All retrieved tables, news, stock prices, textual description of charts and the visual chart itself are used as the input of a multi-modal LLM to derive the answer.  BGE-Visual~\cite{zhou2024vista} is used in this setting.

\header{Evaluation Metrics}
Following the standard evaluation protocol, we use Exact Match (EM), F1 Score, and Accuracy (Acc) as evaluation metrics~\cite{rajpurkar2016squad}.
Additionally, to achieve a comprehensive assessment of model performance, we employ LLMs as automated judges to assess model predictions compared to ground-truth answers.


\header{Implementation Details}
GPT-4o-mini is used to generate the textual description in graph construction, and keywords in retrieval.  
We use text-embedding-3-small to transform text chunks to dense vectors.
GPT-4o-mini is also used as the LLM evaluator.
We use Milvus as the vector database and neo4j as the graph database.
GPT-4o-mini is applied as the multi-modal LLM to take the question and the retrieved results as input to infer the answer.
For BGE-Text and BGE-Visual, we apply bge-large-en-v1.5 and bge-visualized-base-en-v1.5; for CLIP-B and BLIP-B, we use clip-vit-base-patch16 and blip-image-captioning-base.


\begin{table}[t]
\centering
    \setlength{\tabcolsep}{2mm}
\footnotesize
\setlength{\abovecaptionskip}{-0.4pt}
\setlength{\belowcaptionskip}{-6pt}
    \caption{Performance comparison between our \method~and other baseline methods.
    Best and second-best results are marked in bold and underlined, respectively.
    }
    \label{tab:OverallPerformanceComparison}
    \resizebox{\linewidth}{!}{
    \begin{tabular}{llrrrr}
    \toprule
  \bf Setting & \bf Model & \bf EM (\%) & \bf F1 Score & \bf Acc (\%) & \bf LLM Acc (\%) \\ 
    \midrule
     \multirow{1}[0]{*}{No Retrieval} & GPT-4o-mini  & 4.51  & 5.89  & 6.68  & 6.25 \\
    \midrule
    \multirow{6}[0]{*}{No Visual} & BM25  & 10.85 & 20.89 & 15.89 & 11.97 \\
          & Naive RAG & 6.59  & 17.05 & 10.53 & 8.78 \\
          & GraphRAG & 0.05  & 12.86 & 18.57 & 7.01 \\
          & LightRAG & 4.62  & 15.07 & 8.32  & 8.32 \\
          & BGE-Text & \underline{17.11} & \underline{27.36} & \underline{23.41} & \underline{18.50} \\
          & \bf \method & 15.45 & 26.48 & 22.14 & 17.02 \\
    \midrule
    \multirow{2}[0]{*}{All}
          & CLIP-B & 12.12 & 20.30 & 19.75 & 14.33 \\
           & BLIP-B & 13.21 & 22.56 & 20.71 & 15.14 \\
          & BGE-Visual & 14.51 & 25.45 & 21.89 & 16.04 \\   
          & \bf \method & \bf 19.12 & \bf 31.41 & \bf 26.56 & \bf 21.53 \\
    \bottomrule
    \end{tabular}
}
\vspace{-0.6cm}
\end{table}

\begin{figure*}[t]
    \setlength\abovecaptionskip{-0.4px}
    \setlength\belowcaptionskip{-0.6px}
    \centering
    \includegraphics[width=1\linewidth]{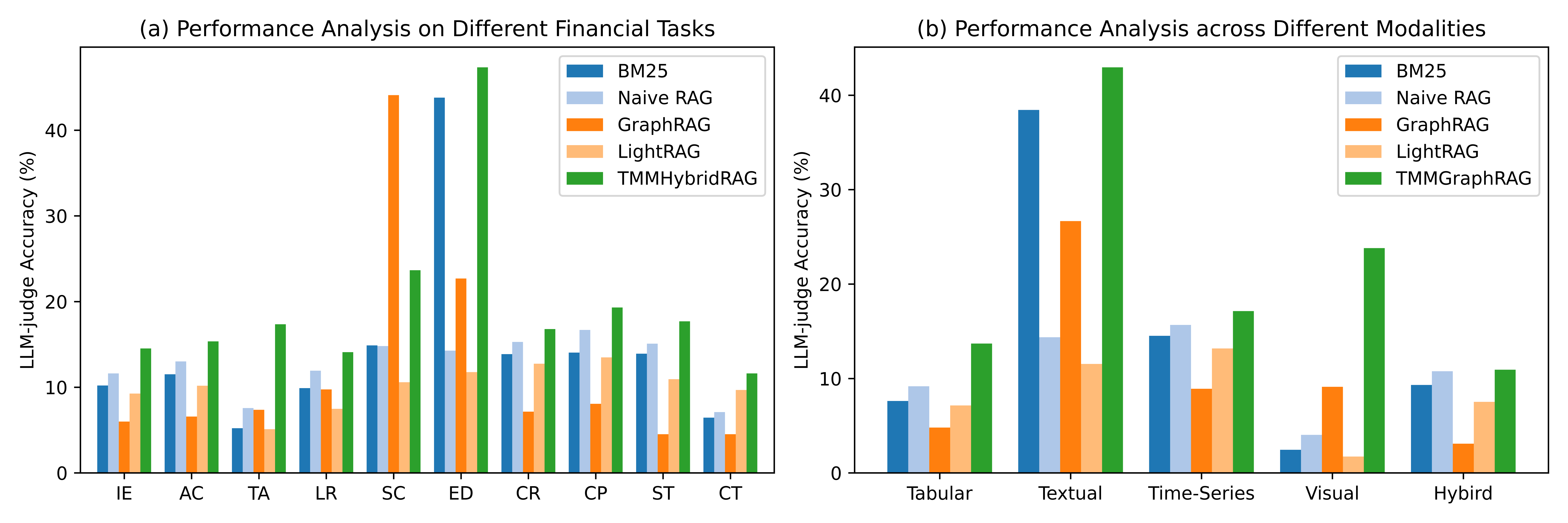}
    \Description{Performance analysis on different financial tasks and modalities. (a) Performance on different financial tasks. (b) Performance on different modalities.}
    \caption{Performance analysis on different financial tasks and modalities.}
    \label{fig:BaselineAccuracy}
    \vspace{-0.3cm}
\end{figure*}

\subsection{Main Results}
To verify the effectiveness of the proposed \method, we compare its performance with baseline methods on the newly constructed \dataset. 
Experiment results are summarized in \autoref{tab:OverallPerformanceComparison}, from which we make several key observations:
    1) Under \emph{No Retrieval} setting, GPT-4o-mini performs poorly, revealing the necessity of the RAG for correctly answering the questions in our \dataset. 
    2) Among all methods in \emph{No Visual} setting, BGE-Text achieves the highest scores compared to other methods. Our \method~(No Visual) ranks the second and reaches comparable performance on all four metrics. 
    3) Our~\method~(All) consistently achieves the best results across all evaluation metrics, demonstrating the superiority of our method in addressing the problems in \dataset. Specifically, it attains an EM score of 19.12\%, an F1 score of 31.41, an accuracy of 26.56\%, and an LLM-judge accuracy of 21.53\%.
    4) Though our \method~(All) achieves state-of-the-art on \dataset, the F1 score remains relatively low at 31.41. This highlights the significant challenges inherent in \dataset, demanding the development of more advanced RAG methods.

\begin{table*}[ht]
    \centering
    \setlength\abovecaptionskip{-0.4px}
    \setlength\belowcaptionskip{-0.6px}
    \setlength{\tabcolsep}{2pt} 
    \begin{threeparttable}
    \caption{Ablation study. Best and second-best results are marked in bold and underlined, respectively.}
    
    \label{tab:ablation_study}
    \begin{tabular}{lrrrrrrrrrrrrrrr}
        \toprule
        \multirow{2}{*}{\textbf{Model}} &  \multirow{2}{*}{\textbf{EM (\%)}} &  \multirow{2}{*}{\textbf{F1 Score}} &  \multirow{2}{*}{\textbf{Acc (\%)}} &  \multirow{2}{*}{ \bf LLM-judge Acc (\%)} & \multicolumn{10}{c}{\textbf{ LLM-judge Acc (\%) on Different Financial Tasks}} \\
        \cmidrule(lr){6-15}
        & & & & & IE & AC & TA & LR & SC & ED & CR & CP & ST & CT \\
        \midrule
        \midrule
        \method~(All)  & 19.12 & 31.41 & 21.53 & 26.56 & 14.19 & 14.57 & 19.90 & 15.23 & 14.74 & 11.38 & 15.86 & 17.02 & 59.77 & 51.01 \\
        \: - Vec & 6.34  & 11.45 & 6.32  & 7.26  & 9.74  & 5.79  & 13.89 & 11.31 & 12.35 & 4.88  & 8.70  & 12.89 & 1.54  & 5.90 \\
        \: - Graph & 14.96 & 25.75 & 16.23 & 20.23 & 12.04 & 12.76 & 14.52 & 13.05 & 14.49 & 9.76  & 13.02 & 14.21 & 36.42 & 44.20 \\
       \: - Raw & 15.63 & 27.58 & 16.75 & 26.44 & 20.38 & 27.40 & 26.03 & 20.55 & 14.98 & 12.20 & 15.47 & 17.75 & 38.41 & 40.44 \\
        \: - Temporal & 17.16 & 28.38 & 19.29 & 23.70  & 12.84 & 10.04 & 18.83 & 14.24 & 13.64 & 13.93 & 12.81 & 16.16 & 56.28 & 50.25 \\
        \bottomrule
    \end{tabular}
    \end{threeparttable}
\vspace{-0.4cm}
\end{table*}




\subsection{In-Depth Analysis}
We further investigate the performance of methods across various financial tasks and data modalities. See results in \autoref{fig:BaselineAccuracy}.

\header{Performance Analysis on Different Financial Tasks} As shown in \autoref{fig:BaselineAccuracy} (a), we can observe: 
1) Our \method~(All) significantly outperforms all other methods on most financial tasks, demonstrating consistent effectiveness across diverse challenges in the finance domain. 
2) For Sentiment Classification which is designed to inquire about specific aspects of a company, requiring the aggregation of dispersed information, GraphRAG achieves the best performance, possibly because its explicit high-level structures, like communities, can particularly benefit the summarization-based reasoning tasks. 
3) For the Event Detection task, BGE-Visual obtains the highest performance, demonstrating the effectiveness of BGE-based models in processing textual news data.

\header{Performance Analysis Across Different Modalities}
We present the performance of all methods across different modalities in \autoref{fig:BaselineAccuracy} (b). We find:
1) \method~(All) consistently beats all other methods across all modalities on our \dataset, underscoring its superiority in answering temporal-aware questions over multi-modal data. 
2) Comparably, 
 \method~is especially effective for questions involving visual technical charts, validating our approach to handling visual data through textual descriptions, temporal information, and raw images.
3) In contrast, questions that rely on multiple modalities and tabular data pose the greatest challenge for the \method~method, highlighting the difficulties of our \dataset. 


\subsection{Ablation Study}
We conduct ablation study to evaluate effects of design choices in \method, including temporal-aware dense vector, temporal-aware heterogeneous graph, raw data mapping, and incorporation of temporal information as properties in entities and relationships.
See experiment results in \autoref{tab:ablation_study}.
\begin{itemize}[leftmargin=*]
\item \header{Removing Temporal-aware Dense Vectors (- Vec)} In this variant, the temporal-aware dense vector is removed. 
Given a query, the model searches for the entities and relationships from the graph only. 
This leads to a significant decline in performance across all four evaluation metrics, e.g. 31.41 down  to 11.45 for F1 score.
The most substantial performance drop is observed on the \emph{Sentiment Classification} and \emph{Event Detection} tasks over news articles.
This reveals the importance of constructing dense vectors for effectively addressing questions that depend on textual data.

\item \header{Removing Temporal-aware Heterogeneous Graph (- Graph)}
This variant removes the temporal-aware heterogeneous graph. 
Given a query, all relevant entities and relationships are retrieved from the temporal-aware dense vectors.
A significant performance drop across all four metrics can be observed.
As \emph{Trend Analysis} requires understanding sequential relationships, the absence of the graph leads to worse performance.
Note, the performance on some tasks, including \emph{Sentiment Analysis}, \emph{Logical Reasoning} and \emph{Counting}, is slightly better than the full model.
This may be because graph retrieval can introduce noise, hindering the multi-modal LLM from identifying correct information.

\item \header{Removing Raw Data Mapping (- Raw)}
This variant chooses not to use raw data during answer generation, relying only on the retrieved entity and their relationships, which leads to a noticeable drop across all metrics. 
For some tasks, e.g. \emph{Arithmetic Calculation} and \emph{Logical Reasoning}, the performance is better than the full model.
This may be because all necessary information for answering the questions is already contained within the entities or relations, and raw data tends to include irrelevant details misleading the multi-modal LLM in answer generation.


\item \header{Removing Temporal Information (- Temporal)} 
This variant removes temporal-related properties from all entities and relations, leading to worse performance than the full model across all four metrics.
The decline is especially obvious on \emph{Arithmetic Calculation} and \emph{Trend Analysis} tasks, highlighting the importance of incorporating temporal information for effectively analyzing temporal-aware calculation and trend analysis in RAG systems.

\end{itemize}

\subsection{Performance Analysis on Different Multi-modal LLMs}
We replace the multi-modal LLM used for answer generation with other multi-modal LLMs and compare their performance.
Compared models are from different model families, 
including GPT-4o-mini~\cite{achiam2023gpt}, Llama 3.2 series~\cite{dubey2024llama}, Qwen series~\cite{Qwen-VL}, DeepSeek series\cite{deepseekai2025deepseekr1incentivizingreasoningcapability}, and Gemini series~\cite{comanici2025gemini}, and Gemini series~\cite{comanici2025gemini}.
In \autoref{tab:llm_comparison} we summarize parameter sizes, multi-modal LLMs, and their corresponding performance on \dataset.
It can be seen that Gemini-2.0-Flash achieves the highest accuracy of 18.42\%, followed by Kimi-VL-A3B-Instruct at 17.63\%, surpassing both DeepSeek series and Qwen series. 
This suggests that even with the closed source models still leading the pack, some open-source models can achieve competitive performance.
It also suggests that model size is not everything, indicating that \method, with its efficient architectures and techniques, does not rely on high-performance multi-modal LLMs to deliver competitive results.
These results further demonstrate the broad applicability and effectiveness of our approach across diverse model classes and settings.

\subsection{Error Analysis}
We analyze error cases to better reveal the limitations of our \method~and the challenges inherent in \dataset.
We randomly select 200 incorrect predictions and categorize the errors into four groups, as shown in \autoref{tab:error-case}, each with a representative example.
1) \emph{Retrieval Error (46.5\%)}: The retrieved data does not contain the key entities, relations, or relevant information needed to answer the question.
2) \emph{Calculation Error(29.0\%)}: The model correctly selects the relevant formula but makes mistakes in computation.
3) \emph{Reasoning Error (13.5\%)}: The model misunderstands financial concepts, misinterprets relationships between variables, or applies incorrect logical reasoning to infer the answer. 
4) \emph{Temporal Error (5.5\%)}: The model uses data from the correct source but associates it with the wrong timestamp.

\begin{table}[t]
\centering
\small
\caption{Performance comparison of different multi-modal LLMs with retrieval.}
\label{tab:llm_comparison}
\begin{tabular}{l l r r}
\toprule
\textbf{Model} & \textbf{Open/Closed} & \textbf{Params (B)} & \textbf{Accuracy (\%)} \\
\midrule
GPT-4o-mini            & Closed-source & --  & 21.53 \\
Gemini-2.0-Flash       & Closed-source & --  & 18.42 \\
Kimi-VL-A3B-Instruct   & Open-source   & 16  & 17.63 \\
Qwen2.5-7B-Instruct    & Open-source   & 7   & 15.15 \\
DeepSeek R1 8B         & Open-source   & 8   & 13.41 \\
DeepSeek R1 14B        & Open-source   & 14  & 13.15 \\
Llama 3.2 11B          & Open-source   & 11  & 11.57 \\
Llama 3.2 3B           & Open-source   & 3   & 11.27 \\
Qwen-VL-Chat           & Open-source   & 7   & 8.89  \\
\bottomrule
\end{tabular}
\vspace{-0.8cm}
\end{table}
We make following observations:
1) Most errors are \emph{Retrieval Errors} (46.5\%). 
This suggests advanced indexing or retrieval methods are demanded to improve recall in information retrieval.
2) Challenges persist in Arithmetic Computation and Complex Reasoning.
\emph{Calculation Errors} and \emph{Reasoning Errors} collectively account for 42.5\% of failures, underscoring the challenges multi-modal LLMs face in performing arithmetic computations and complex reasoning. 

To address these issues, two possible approaches can be considered:
i) to improve quality of retained information in the retrieval, such as reducing irrelevant content;
ii) to enhance LLMs' understanding of financial terminology, improve their ability to perform complex financial reasoning, and integrate external tools to assist with numerical computations.
iii) Temporal Inference is crucial.
Though less frequent, \emph{Temporal Errors} (5.5\%) are unignorable for time-sensitive tasks, as incorrect temporal inference can result in significant factual inaccuracies. 

\section{Related Work}
\subsection{Financial QA Datasets}
To date, many financial QA datasets have been released to advance research in financial analysis, which can be divided to \emph{Non-RAG QA}, \emph{Text-RAG QA}, and \emph{Multi-Modal-RAG QA} datasets.
\emph{Non-RAG QA}~\cite{maia201818,zhu2021tat, xie2024finben} datasets focus on financial analysis using relatively short context information that can be directly input into LLMs.
For example, FiQA-SA \cite{maia201818} and FPB \cite{malo2014good} are designed for emotion analysis based on financial texts;
TAT-QA~\cite{zhu2021tat} and FinQA~\cite{chen2021finqa} aim to answer questions given a financial table and its associated paragraphs extracted from financial reports.
\emph{Text-RAG QA} datasets, e.g.  FinTextQA \cite{chen2024fintextqa} and OmniEval \cite{wang2024omnieval},
are aimed at evaluating text-based RAG systems in finance. 
For instance, FinTextQA~\cite{chen2024fintextqa} is a long-form QA dataset containing 1,262 high-quality QA pairs that require RAG systems to address based on finance textbooks and policy and regulation from government agency websites.
Current \emph{Multi-Modal RAG QA} datasets include FinanceBench~\cite{islam2023financebench}, incorporating time-series data in addition to textual data, and AlphaFin~\cite{li2024alphafin}, involving visual data with textual data to assess RAG systems.
Though with notable strengths, these datasets are limited to specific modalities, and only AlphaFin incorporates some temporal questions focused on time-series data. 
In comparison, our \dataset~is the first temporal-aware multi-modal benchmark designed to evaluate RAG systems in finance. 
It encompasses financial data across four modalities—tabular, textual, time-series, and visual data. Additionally, all questions in \dataset~are temporal-aware, addressing a critical gap in existing benchmarks.

\begin{table}[t]
    \setlength\abovecaptionskip{-0.3px}
    \setlength\belowcaptionskip{-0.3px}
    \caption{Error Analysis. Q, G, P denote question, golden answer, and~\method~generated answer, respectively.
     }
    \label{tab:error-case} 
    \centering
    \footnotesize
    \renewcommand{\arraystretch}{1.2} 
    \setlength{\tabcolsep}{2pt} 
    \begin{tabular}{p{0.15\textwidth}|p{0.3\textwidth}}
        \toprule
        \multirow{5}{*}{\makecell[l]{Retrieval Error \\ (46.5\%)}} 
        & \textbf{Q}: What was CoStar Group's \textit{otherCurrentAssets} value on March 31, 2022? \\
        & \textbf{G}: \textcolor{blue}{USD 36,183,000} \\
        & \textbf{P}: The retrieved tables \textcolor{red}{do not contain} any data the \textit{otherCurrentAssets} value. \\
        \midrule
        \multirow{6}{*}{\makecell[l]{Calculation Error \\ (29.0\%)}} 
        & \textbf{Q}: If Datadog had 15,000,000 shares instead of 10,000,000 and a book value of USD 2,000,000,000 , what would its P/B ratio be on Jan 5, 2022? \\
        & \textbf{G}: \textcolor{blue}{1.036} \\
        & \textbf{P}: Book Value per Share: \textcolor{red}{$\frac{2,000,000,000}{10,000,000} = 20$}\\
        \midrule
        \multirow{6}{*}{\makecell[l]{Reasoning Error \\ (13.5\%)}} 
        & \textbf{Q}: If Ansys's stock price trend from October 13, 2022, continued, what would its price be next month? \\
        & \textbf{G}: 207.68 * (1 + 0.0769) = \textcolor{blue}{USD 223.66 } \\
        & \textbf{P}: With the price reaching \textcolor{red}{a last closing price of USD 279.21 }... \\
       
        \midrule
        \multirow{4}{*}{\makecell[l]{Temporal Error \\ (5.5\%)}} 
        & \textbf{Q}: When did AEP experience the lowest price in September 2022? \\
        & \textbf{G}: \textcolor{blue}{September 30, 2022} \\
        & \textbf{P}: \textcolor{red}{On October 29, 2022}, the stock ...\\
        \bottomrule
    \end{tabular}
    \vspace{-0.4cm}
\end{table}
 
\subsection{Graph-based RAG}

RAG \cite{Lewis2020RAG,zhu2021retrieving} has been widely used to enhance performance of LLMs across various tasks by integrating an Information Retriever (IR) module to leverage external knowledge. Recently, graph-based RAG methods~\cite{edge2024local,hybridrag,guo2024lightrag,zhu2023soargraph,zhu2023doc2soargraph} have demonstrated remarkable performance across diverse applications. 
For instance, GraphRAG~\cite{edge2024local} improves traditional RAG by building a knowledge graph from extracted entities and relations, grouping related entities into communities, and generating summaries for each. During inference, it synthesizes answers from these community summaries.
Hybrid~\cite{hybridrag} and LightRAG~\cite{guo2024lightrag} enhance GraphRAG by combining dense retrieval with graph retrieval techniques. Despite effectiveness, all these methods primarily focus on textual data, resulting in suboptimal performance when handling multi-modal data. Moreover, they struggle to effectively address temporal-aware queries in \dataset. 
We propose \method, a novel graph-based RAG approach specifically designed to tackle the challenges of temporal-aware multi-modal RAG presented in \dataset.

\section{Conclusion}

In this work, we introduce \dataset, the first benchmark for evaluating temporal-aware multi-modal Retrieval-Augmented Generation (RAG) systems in financial analysis. \dataset~comprises 5,676 questions spanning financial tables, news articles, stock prices, and technical charts, designed to assess a model’s ability to retrieve and reason over temporal financial information. To address its challenges, we propose \method, a novel approach integrating dense and graph retrieval with temporal-aware entity modeling. Our experiments show \method~ outperforms existing methods, yet the generally low performance also highlights the persisting challenges of our \dataset. 


\bibliographystyle{ACM-Reference-Format}
\bibliography{sample-base}

\end{document}